# Principle of Relativity in Physics and in Epistemology

Guang-jiong Ni

Department of Physics, Fudan University, Shanghai 200433, China
Department of Physics, Portland State University, Portland, OR97207, USA
(Email: pdx01018@pdx.edu)

Abstract: The conceptual evolution of principle of relativity in the theory of special relativity is discussed in detail . It is intimately related to a series of difficulties in quantum mechanics, relativistic quantum mechanics and quantum field theory as well as to new predictions about antigravity and tachyonic neutrinos etc.

## I. Introduction: Two Postulates of Einstein

As is well known, the theory of special relativity(SR) was established by Einstein in 1905 on two postulates (see,e.g.,[1]):
Postulate 1: All inertial frames are equivalent with respect to all the laws of physics.
Postulate 2: The speed of light in empty space always has the same value c.

The postulate 1 , usually called as the "principle of relativity" , was accepted by all physicists in 1905 without suspicion whereas the postulate 2 aroused a great surprise among many physicists at that time.

The surprise was inevitable and even necessary since SR is a totally new theory out broken from the classical physics. Both postulates are relativistic in essence and intimately related. This is because a law of physics is expressed by certain equation. When we compare its form from one inertial frame to another, a coordinate transformation is needed to check if its form remains invariant. And this Lorentz transformation must be established by the postulate 2 before postulate 1 can have quantitative meaning. Hence, as a metaphor, to propose postulate 1 was just like " to paint a dragon on the wall" and Einstein brought it to life "by putting in the pupils of its eyes"(before the dragon could fly out of the wall) via the postulate 2[2].

The above understanding explains why all attempts at deriving SR by postulate 1 solely or by weakening postulate 2 to some extent were doomed to failure. However , to combine these two postulates into one is possible only after we can refine the whole achievements of physics in the 20$^{th}$ century and the refinement is possible only after we gradually understood the deep meaning of SR in the past 100 years. The purpose of this paper is focused on the principle of relativity being not only a principle in physics , but also a principle in epistemology. These two aspects, however, are indivisible and comprise the main clue to understand many difficulties in quantum mechanics(QM) , relativistic QM (RQM), quantum field theory(QFT), particle physics and cosmology as well.

## II. Lorentz Transformation, Minkowski Diagram and Poincare Group

Einstein historically and correctly evaluated the great achievement of Newton by a remark that J. Wheeler often remembered and repeated, how much Einstein admired Newton's courage and judgment in sticking with the idea of absolute space and time even though all Newton's colleagues told Newton it was absurd. This is because absolute space and time was what was required to make progress at Newton's time ([3],p.146). And Einstein clearly knew in 1905 that it was the time to abandon the concept of absolute space and time because " the justification for a physical concept lies exclusively in its clear and unambiguous relation to facts that can be experienced"([1],p.70).

In particular, based on his postulate 2, Einstein analyzed the relativity of simultaneity. He pointed out that an observer A (or B) resting on an inertial frame S(or S') must have his own clock to record the time t (or t') for a same "event". In contrast to t=t' in Newtonian mechanics, now in SR they are not only different, but also linked to spatial coordinate x (or x') at which the event occurs:

$$x' = \frac{x - vt}{\sqrt{1 - v^2/c^2}} \quad , \quad t' = \frac{t - vx/c^2}{\sqrt{1 - v^2/c^2}} \qquad (2.1)$$

where v is the relative velocity between frames S and S' along the x(x') axis. The symmetry between S and S' can be seen from the inverse transformation :

$$x = \frac{x' + vt'}{\sqrt{1 - v^2/c^2}} \quad , \quad t = \frac{t' + vx'/c^2}{\sqrt{1 - v^2/c^2}} \qquad (2.2)$$

Eqs. (2.1) and (2.2), together with y=y', z=z', are called the Lorentz transformation, which was first written down by Lorentz in 1904. But Einstein discovered it independently a year later and gave it a totally new and correct explanation.

In 1908, Einstein's teacher, a mathematician H. Minkowski put Eq. (2.1) into a diagram with x and t being the abscissa and ordinate of a (rectangular) S frame. Then x' and t' will become nonrectangular coordinates of S' frame on that of S frame (see e.g.,p.83 of [1]). The important thing is: although an event P has different coordinates (x, y, z, t) and (x', y', z', t') in S and S' frames, there is a space-time invariant:

$$c^2 t^2 - (x^2 + y^2 + z^2) = c^2 t'^2 - (x'^2 + y'^2 + z'^2) = \text{constant} \qquad (2.3)$$

In the group theory of mathematics, all possible transformations capable of preserving some invariant will form a group. For example, the rotational group in three dimensional Euclidean space is formed by all possible rotations to keep the three dimensional distance D invariant:

$$x^2 + y^2 + z^2 = x'^2 + y'^2 + z'^2 = D^2 \qquad (2.4)$$

So Lorentz transformations, (2.1),(2.2) etc, denoted by $x'^{\mu} = \tilde{U}^{\mu}_{\nu} x^{\nu}$ form a Lorentz group with the rotational group in three dimensional space a subgroup of it.

Every finite transformation can be reached by the accumulation of infinite infinitesimal transformations. In other words, it can be achieved by continuous actions of (infinitesimal) generators. For example, the rotational group has three generators which just correspond to three angular momentum operators in QM[2]. Of course, Lorentz group has extra three generators. Moreover, if taking all inhomogeneous Lorentz transformations

$$x'^{\mu} = \tilde{U}^{\mu}_{\nu} x^{\nu} + a^{\mu} \tag{2.5}$$

into account with $a^{\mu}$ denoting a constant translation along $x^{\mu}$, the Poncare group is formed [4]. It has another four generators $P_{\mu}$ (of the space-time translation group), which will be discussed further in the next section.

According to the theory of Poincare group, there are two Casimir invariants commuting with all ten generators. One of them is simply $P_{\mu} P^{\mu}$ with eigen values $m^2$ (m being the mass of a particle), i.e.,:

$$P_{\mu} P^{\mu} = m^2 \tag{2.6}$$

We will see later that both Minkowski's point of view (which was highly admired by Einstein) and Eq. (2.6) are very important to physics.

### III. Observables and Nonobservables

It was often said that SR and QM are two great pillars of modern physics. Undoubtly, they are equally important. What will be discussed in this paper is: SR is not only the promoter of QM, but also an interpreter of QM. We will discuss the former aspect in this and next two sections while the latter one in section VII.

In 1926, Einstein attended a lecture on QM (then called the matrix mechanics) by Heisenberg in Berlin University. On the way home, they had a conversation as follows:[5]

Einstein said:" You ignored the orbital motion of electron inside an atom. But the electron trajectory can be clearly seen , say, in a cloud chamber. Do you still refuse to consider it ?" He asked again:" Do you truly believe that one can build physical theory solely based on observables?" Heisenberg was astonished and said:" Sir, I just learnt from you. Didn't you insist on doing so in the SR? You taught us not to discuss the absolute time but the relative time and the reason why the absolute time is meaningless is just because it is unobservable." Then Einstein replied:" …In principle, we cannot construct the theory solely based on observables. Actually, it is the theory which determines what we can observe."

The above conversation between Einstein and Heisenberg is really very interesting and important. They pushed the principle of relativity in SR into a deeper level of epistemology . What essential in their discussion is: an absolute thing is unobservable, but we still need nonobservables to establish a theory which in turn tells us what will be seen in real experiments.

The development of physics in the 20[th] century reveals that a transformation of nonobservable will lead to a relevant conservation law (or selection rule) for an observable as summarized by T.D.Lee [6]. For instance, the absolute time t (or absolute position in space, x) is unobservable, so we can perform a time (space) translation:

$$x^\mu \to x^\mu + a^\mu \tag{3.1}$$

at our disposal without changing the law of physics. This invariance under the time (space) translation leads to the energy E (momentum p) conservation law. Both E and p are observables. In QM, their relations are expressed ( in one dimensional space only)as:

$$\hat{p} = -i\hbar \frac{\partial}{\partial x}, \qquad \hat{E} = i\hbar \frac{\partial}{\partial t} \tag{3.2}$$

which are just the corresponding generators of spase-time translational group with Plank constant $\hbar$ as their common multipliers. The deep implication of (3.2) lies in the fact that coordinates x and t in the wave function(WF) of QM are not direct observables and the momentum p and energy E are observable only in the sense of operators, which means that p and E are unobservable until some action is performed on the quantum system(say, a particle) . In other words, one has to push the particle along the x axis (or along t) a little bit before the information of p (or E) can be created. In short , no change, no information. This is just because an absolute thing is transformed into a relative thing only during the changing process.

Even Dirac, one of the great founders of QM, had at one time (1927) overlooked the above principle. He introduced a Hermitian operator $\hat{\phi}$ for the "phase" of photon field and established its commutation relation with the photon number operator $\hat{N}$ as:

$$[\hat{\phi}, \hat{N}] = -i \tag{3.3}$$

But it is essentially wrong. Just like the absolute time t, an absolute phase (angle) cannot be a Hermitian operator corresponding to an observable. Only a relative angle, i.e., a phase difference, can be observable. Then two (not one) Hermitian operators for phase were introduced correctly by Susskind and Glogower(see [7] and section 1,6D in [2]), just like a motto in Chinese philosophy saying: "The 'one' is invisible until 'two' are established".

## IV. Negative Energy and Antiparticle

In his second short paper establishing SR, Einstein proposed the greatest equation changing the world:

$$E = mc^2 \tag{4.1}$$

which has been verified by all experiments. Here both energy E and mass m are measured to be positive.

However, in 1928, aiming at combining SR with QM, Dirac discovered an equation for electron, which not only explained the electron spin , but also predicted the existence of its antiparticle , the positron ( discovered by Anderson in 1932) ,by showing that the energy E can be negative ,e.g., in the case of free motion:

$$E = \pm\sqrt{p^2 c^2 + m_o^2 c^4} \tag{4.2}$$

where $m_o$ is the rest mass of particle. Similar negative energy was also found in the Klein-Gordon (KG) equation for spinless particles. The reason why the negative energy emerges in QM is because in the WF of particle:

$$\psi(x,t) \approx \exp[i(px - Et)/\hbar] \tag{4.3}$$

only one set of operator formula (3.2) was used to get (4.2). In order to explain the existence of positron with positive energy, many physicists resorted to a "hole theory" based on the "sea" filling up by "negative-energy electrons", which actually has no guarantee from the Pauli exclusion principle, let alone the KG particles. Only a few physicists, like Schwinger[8], Konopinski and Mahmaud[9], even earlier but implicitly, Feynman[10] and Stueckelberg[11], pointed out the necessity of using:

$$\hat{p}_c = i\hbar \frac{\partial}{\partial x} \quad , \quad \hat{E}_c = -i\hbar \frac{\partial}{\partial t} \tag{4.4}$$

for antiparticles (denoted by subscript c ) as the counterpart of (3.2) for particles. Eq. (4.4) means that once E<0 in (4.3), we should write the WF for an antiparticle as:

$$\psi_c(x,t) \sim \exp[-i(px - Et)/\hbar] \tag{4.5}$$

In a long review article by Lee and Wu[12], they defined correctly the antiparticle state $|\bar{a}\rangle$ being the CPT transformation of its particle state $|a\rangle$:

$$|\bar{a}\rangle = CPT |a\rangle \tag{4.6}$$

to replace the original one

$$|\bar{a}\rangle = C |a\rangle \tag{4.7}$$

which was already proven to be wrong after the discovery of parity (P) violation in weak interactions in 1956-1957. Expressing (4.6) in terms of WF, we easily obtain (4,5) for either Dirac or KG particles (see [13] apart from some irrelevant matrix prefactor).

Some remarks are in order:

(a). Why (4.7) is wrong ? Many physicists believed that the distinction between electron and positron is absolute----they carry opposite " intrinsic charge" -e and e (e>0) and there is a " conservation law of charge q" expressed as a continuity equation which was used by Maxwell to introduce the "displacement current" for establishing the theory of classical electrodynamics(CED) . However, the conservation law of charge is no longer valid since the experimental discovery of the variation of fine-structure constant $\alpha = e^2/\hbar c$ increasing from 1/137 at low energy to 1/128 at high energy ( for theoretical explanation ,see [14]) . And the continuity equation in CED is already substituted by its counterpart in QM without the appearance of a charge unit e[2]. So the conservation law of charge is already substituted by that of charge number $Q = q/e$, e.g., $Q = -1$ and 1 for electron and positron as reflected in the distinction between i and -i in the WF (4.3) versus (4.5). The distinction between them is merely relative in the phase of space-time evolution, not an absolute one in some " intrinsic space".

(b). The definition of so-called time-reversal---T transformation reads [15,2]:

$$\psi(x,t) \to \psi^*(x,-t) \tag{4.8}$$

which means

$$\exp[i(px - Et)/\hbar] \to \exp[i(-px - Et)/\hbar] \tag{4.9}$$

by Eq. (4.3), just implying a reversal of motion. Hence the name "time reversal (T)" is indeed a misnomer and its tiny violation (i.e., CP violation) discovered in weak interactions since 1964 has nothing to do with the symmetry of pure time inversion ($t \to -t$) without complex conjugation,( which is violated to 100% in nonrelativistic QM) or the symmetry between particle and antiparticle. The latter symmetry is ensured by the CPT theorem as shown in (4, 6). But actually, the CPT theorem already becomes a postulate which can be stated as follows:

Under the (newly defined ) space-time inversion $\mathscr{PT}$ ($x \to -x, t \to -t$), the theory of relativistic QM (RQM) remains invariant while a concrete state of particle transforms into that of its antiparticle :

$$\mathscr{PT} = \mathscr{C} \qquad (4.10)$$

where $\mathscr{C}$ means the particle –antiparticle transformation , whose definition is contained in the postulate (4.10), not coming from else where.

(c). The above situation bears some resemblance to what happened in classical physics. The definition of inertial mass of a body, m, is just given by the ratio of force F to acceleration a , i.e., by the Newton's law:

$$F = m a \qquad (4.11)$$

Similarly, the definition of electric (magnetic) field strength $\vec{E}(\vec{B})$ is contained in the law of Lorentz force:

$$\vec{F} = m\vec{a} = q(\vec{E} + \vec{v} \times \vec{B}/c) \qquad (4.12)$$

A law (postulate) can often ( not always) accommodate a definition of physical quantity and the validity of law together with the definition has to be verified by experiments, not by the theory coming from else where. This is the method of induction. By contrast, all quantities in a theorem must be defined independently in advance and the outcome deduced from the theorem is actually hidden in its premise. However, although CPT theorem has been tested by experiments, once the conservation laws of C, P, and T are discovered to be violated individually, their definitions cease to be meaningful in physics. We should change the definition of space-time inversion into $\mathscr{PT}$ and regard the particle-antiparticle transformation $\mathscr{C}$ not an independent one but a direct consequence of $\mathscr{PT}$ as shown in (4.10). The relation between (4.3) and (4.5), or that between (3.2) and (4.4), is just what (4.10) means. Eq.(4.10) is also a starting point of QFT, e.g., it gives the exact definition of a field operator [2,36].We have been waiting for the conception change for too long a time.

## V. Negative mass paradox, Invariance of mass inversion and Antigravity

If we insist on Eq. (4.1) and emphasize the inevitable appearance of negative energy in RQM, we must consider the negative mass as well. Interesting enough, a possible negative gravitational mass had been discussed by Bondi[16], Schiff[17] and Will[18]. Suppose that a particle with negative mass $m_1 < 0$ is brought close to another particle with positive mass $m_2 > 0$. Then according to Newton's gravitation law:

$$F(r) = -G\frac{m_1 m_2}{r^2} \qquad (5.1)$$

and Eq. (4.11), $m_2$ would attract $m_1$ whereas $m_1$ would repel $m_2$. The pair ( a "gravitational dipole") would accelerate automatically in space, a bizarre phenomenon that no one can believe and so it could be named as a "negative mass paradox" in gravity[19].

To get rid of this paradox once and for all, we should notice that in previous sections the distinction between (x, t) and ( - x, -t ) , or that between E and – E , is merely relative in the sense of symmetry transformation to show the equal existence of particle versus antiparticle. Eventually, there is no negative energy at all. Hence, to keep (4.1) intact, we must treat the mass inversion m → -m also as a symmetry transformation to reflect the particle-antiparticle symmetry and eventually , there is no negative mass at all . Indeed, the theory of RQM does remain invariant under the mass inversion and it can be generalized to a classical theory. For example, a mass inversion (replacing the transformation q → – q) performed on Eq.(4.12) can bring the equation for electron into that for positron. Similarly, Eq. (5.1) should be generalized to:

$$F(r) = \pm G\frac{m_1 m_2}{r^2} \qquad (5.2)$$

so that it can remain invariant under the mass inversion ($m_i \to -m_i, i = 1,2$). Eq. (5.2) implies attractive force between matters (antimatters) whereas repulsive force between matter and antimatter.

Of course, Eq. (5.2) is still a postulate or conjecture and should be subject to tests in cosmology. But it seems much simpler than the hypothesis of so-called "dark energy" to account for the acceleration of universe expansion.[19].

## VI. Parity Violation and a Minimal Three-Flavor Model for Tachyonic Neutrino

Since the discovery of parity violation by Lee-Yang[20] and Wu et al.[21] in 1956-1957 as well as the genius experiment by Goldhaber et al.[22], the permanent helicity of neutrino (antineutrino) being left-handed ( right-handed) has been verified beyond any doubt, especially because it was assumed that neutrinos have zero mass and move at a speed of light c. However, the experimental verification of neutrino oscillation (among three flavors $\nu_e, \nu_\mu$ and $\nu_\tau$ ) in 1998 [23] and the accurate data on solar neutrinos in 2002[24] had excluded any possibility of a massless neutrino. Meanwhile, the measurements on tritium beta decay gave a remarkable result that in the kinematical relation of neutrino [25]:

$$E^2 = p^2 c^2 + m_\nu^2 c^4$$

the mass square seems negative:

$$m^2(\nu_e) = -2.5 + 3.3 eV^2 \qquad (6.1)$$

Evidently, if neutrino is a subluminal particle with nonzero rest mass , no simple theory can be constructed to account for all experimental facts available. However, since 1980's, some authors have been considering an interesting and simple possibility that neutrinos might be superluminal particles (tachyons) with nonzero tachyon mass $m_s$ (real and positive) which can be seen as an analytical continuation of a rest mass $m_0$ ( for subluminal particles):

$$m_0^2 \to -m_s^2, \quad m_0 \to i\, m_s \tag{6.2}$$

In ref. [26], a minimal three-flavor model for tachyonic neutrino is proposed by an equation ($\hbar = c = 1$, $\vec{\sigma}$ are Pauli matrices):

$$\begin{cases} i\dot{\xi}_e = i\vec{\sigma}\cdot\tilde{\mathbf{N}}\xi_e - \delta(\eta_\mu + \eta_\tau) \\ i\dot{\eta}_e = -i\vec{\sigma}\cdot\tilde{\mathbf{N}}\eta_e + \delta(\xi_\mu + \xi_\tau) \\ i\dot{\xi}_\mu = i\vec{\sigma}\cdot\tilde{\mathbf{N}}\xi_\mu - \delta(\eta_\tau + \eta_e) \\ i\dot{\eta}_\mu = -i\vec{\sigma}\cdot\tilde{\mathbf{N}}\eta_\mu + \delta(\xi_\tau + \xi_e) \\ i\dot{\xi}_\tau = i\vec{\sigma}\cdot\tilde{\mathbf{N}}\xi_\tau - \delta(\eta_e + \eta_\mu) \\ i\dot{\eta}_\tau = -i\vec{\sigma}\cdot\tilde{\mathbf{N}}\eta_\tau + \delta(\xi_e + \xi_\mu) \end{cases} \tag{6.3}$$

containing only one coupling parameter $\delta$, which in turn leads to eigenvalues of energy square being:

$$E_j^2 = p^2 - m_j^2 \quad (j=1,2,3)$$
$$m_1^2 = 4\delta^2, \quad m_2^2 = m_3^2 = \delta^2 \tag{6.4}$$

The particle velocity $u_j$ equals to the group velocity $u_g$ of wave:

$$u_j = u_g = \frac{dE_j}{dp} = pc^2/E_j > c \tag{6.5}$$

which is indeed exceeding the light speed c whereas the phase velocity $u_p$ of wave is:

$$u_p = \frac{E_j}{p} = c^2/u_g < c \tag{6.6}$$

If the left-handed chiral states $\xi_i$ ($i = e, \mu, \tau$) dominate the right-handed ones $\eta_i$, i.e., $|\xi_i| > |\eta_i|$, the solution of (6.3) reads:

$$\xi_i \sim \eta_i \sim \exp[i(px - Et)/\hbar] \quad (E > 0) \tag{6.7}$$

describing a neutrino with permanent helicity $<\vec{s}\cdot\hat{p}> = -1$ ($\hat{p}$ means unit vector along $\vec{p}$). Alternatively, if $\eta_i$ dominate $\xi_i$, $|\eta_i| > |\xi_i|$, then we have:

$$\eta_i \sim \xi_i \sim \exp[-i(px - Et)/\hbar] \quad (E > 0) \tag{6.8}$$

describing an antineutrino with helicity 1. It seems that the model (6.3) is capable of explaining the Goldhaber experiment, the coexistence of the (tachyon) mass with parity violation and the equal population among three flavors in the solar neutrino experimental data [24], etc. Moreover, it may explain the appearance of so-called two "knees" in the cosmic ray spectrum located at around proton energy $E = 10^{15.5}$ eV and $10^{17.8}$ eV by the sole parameter $\delta = 0.34$ eV [27]. We assume that an exotic realization of invariance of Eq. (6.3) under the pure time inversion (x → x, t → -t) with

$$\xi_i(x,-t) \to \eta_i(x,t), \quad \eta_i(x,-t) \to \xi_i(x,t) \tag{6.9}$$

will transform a neutrino into an antineutrino in the rest frame of proton, opening suddenly a reaction channel of $\bar{\nu}_e + p \rightarrow n + e^+$ and $\bar{\nu}_\mu + p \rightarrow \Lambda + \mu^+$ at the energy of two knees respectively. For further discussion, see next section.

## VII. Implication of Wave-function in Quantum Mechanics

Despite the great success of SR, QM and quantum field theory (QFT), they are still not so satisfying. The situation was described by Dirac as follows[28]:

" It would seem that we have followed as far as possible the path of logical development of the idea of QM as they are at present understood. The difficulties, being of a profound character, can be removed only by some drastic change in the foundations of the theory, probably a change as drastic as the passage from Bohr's orbit theory to the present QM."

In 1964, in his lecture at Cornell University, Feynman said:"…I think I can safely say that nobody understands QM…." [29].

Why QM is so difficult to understand? Where is the clue?

Dirac had said that before 1972 he thought the commutation relation in QM being the most subtle thing. After 1972, he changed his mind and thought that it is the wavefunction (WF) being the most subtle thing which contains i and thus is unobservable.

We agree with Dirac's opinion that i enters QM essentially and the main difficulties of QM are focused on WF. Let us look at WFs (4.3) vs. (4.5) [or (6.7) vs. (6.8)], there are " observables" p and E hiding in unobservable WFs. What do they mean?

First of all, we should emphasize a radical difference between eastern and western philosophy. In western world, the main stream philosophy was represented by Democritus (460-370 BC). He said :" Nothing exists except atoms and empty space, every thing else is opinion." [30]. His thought played an important role in promoting the science progress in the past 500 years. In the same period, eastern philosophy posed nearly no considerate influence on the science progress because it didn't address the right problem at the right time. However, it's time for eastern philosophy to contribute something meaningful to physics.

In the opening line of his " Tao Te Ching ", Chinese philosopher Lao Tzu said :" The Tao that can be expressed is not the eternal Tao. The name that can be named is not the permanent name."[31] Here the "Tao" means 'way', 'reality' or 'law' and the name that can be named refers to a concrete thing that can be put into certain category. Similarly, it was quoted from the Indian philosophy book Upanishads that[31]:

> There the eye goes not,
> Speech goes not, nor the mind.
> We know not , we understand not.
> How one would teach it?

Combing the above idea with Kant's philosophy in the end of 18$^{th}$ century, we may say that an object remains as a "thing in itself" which contains no information until it is observed and transformed into "thing for us " which is often called the phenomena containing various information ( position x, momentum p and energy E ,etc.) A measurement on the object implies an operation, i.e., a changing on it. The information is created just during the changing, not before. This is why a physical quantity ( say, p ) in classical physics becomes an operator (say, $-i\hbar \partial/\partial x$) in QM.

Then how can we predict what will happen before we perform a real measurement on an object? It is just at this crucial point, the principle of relativity of SR evolves into a principle of relativity in epistemology which can be stated as follows:

A thing can only be cognized during its motion and change relative to others. If isolated from its opposite, it would be bound to become absolute and mysterious object devoid of any understanding.

Say, for a particle, what are its opposites? Among them (the environment, apparatus, etc), the most important one is just the subject---we human being as the observer. Following Dirac, we denote the abstract particle state by $|y\rangle$ which has no representation (no information). If we wish to predict the outcome of a position x measurement, we introduce a basis vector $|x,t\rangle$ (in Heisenberg picture) and get the WF as:

$$y(x,t) = \langle x,t|y\rangle \tag{7.1}$$

Physically, we may understand $|x,t\rangle$ as a "fictitious apparatus" for measuring the position of particle at point x and time t. Both x and t belong to "apparatus"(not to particle). Then why WF (7.1) is expressed as a field like (4.3) or (4.5):

$$\exp[\pm i(px - Et)/\hbar] = \cos(kx - wt) \pm i\sin(kx - wt) \tag{7.2}$$

At this next crucial point, the philosophy initiated by Lao Tzu and Greek philosopher Herakleitos (~500 BC) came to help. Essentially, every thing is by no means composed from indivisible particles but always contains two opposite sides as shown via the real and imaginary parts (both real) divided by i. However, we have to be cautious not to regard the distinction of these two sides (called "ying" and "yang" in Chinese philosophy) being realistic or absolute. Their distinction is merely abstract and relative. One may perform a phase transformation on the WF to see ying and yang changing each other immediately at one's disposal. However, the distinction is sharp enough so that a relative sign change between them in (7.2) amounts to a change from particle to antiparticle.

Therefore, to simulate an interaction (contradiction) between the object $|y\rangle$ and the "fictitious apparatus"$|x,t\rangle$, the WF reflects a possibility of "fictitious measurement". Following Herbert's suggestion that the WF is a "possibility wave" [32] rather than "matter wave", we can understand what Born's probability interpretation of WF means is:

$$\text{real probability} = |WF|^2 = |potential\ possibility|^2 \tag{7.3}$$

The word "potential" is used to stress the invisibility of WF, also the flexibility of the choice of basis vector. For example, we can choose another "fictitious apparatus for measuring the momentum p", $|p,t\rangle$, to express the WF of $|y\rangle$ as:

$$f(p,t) = \langle p,t|y\rangle \tag{7.4}$$

For a same $|y\rangle$, we can find several WFs with different representations. Similar experience was familiar to mathematicians like Minkowski as discussed in section II. An elementary example is an abstract vector in Euclidean space, $\vec{V}$, which represents a geometrical object without any number attached to it. In order to describe it concretely, mathematicians introduce different coordinate frames with unit basis vectors $\vec{e}_i\ (i = x, y, z)$ and $\vec{e}'_i$, yielding:

$$\vec{V} = V_x \vec{e}_x + V_y \vec{e}_y + V_z \vec{e}_z = V'_x \vec{e}'_x + V'_y \vec{e}'_y + V'_z \vec{e}'_z$$

where $\quad V_i = \vec{e}_i \cdot \vec{V}\ ,\ V'_i = \vec{e}'_i \cdot \vec{V}$ (7.5)

are different numbers to represent a same vector.

We should notice the resemblance of (7.1) and (7.4) to (7.5) and learn from mathematicians that the existence of an object and its representation are linked together but they are two different things.

Einstein was correct to say that "it is the theory which determines what we can observe". Let us write the Dirac equation in two –component spinor form:

$$\begin{cases} i\hbar \dot{\boldsymbol{j}}_D = ic\hbar \vec{\boldsymbol{s}} \cdot \nabla \boldsymbol{c}_D + m_o c^2 \boldsymbol{j}_D \\ i\hbar \dot{\boldsymbol{c}}_D = ic\hbar \vec{\boldsymbol{s}} \cdot \nabla \boldsymbol{j}_D - m_o c^2 \boldsymbol{c}_D \end{cases} \quad (7.6)$$

instead of the covariant form in which the WF of electron $\boldsymbol{y} = \begin{pmatrix} \boldsymbol{j}_D \\ \boldsymbol{c}_D \end{pmatrix}$ is written in four-component form. Eq. (7.6) is invariant under the (newly defined) space-time inversion ($x \to -x,\ t \to -t$) with transformation:

$$\boldsymbol{j}_D(-x,-t) \to \boldsymbol{c}_D(x,t)\ ,\quad \boldsymbol{c}_D(-x,-t) \to \boldsymbol{j}_D(x,t) \quad (7.7)$$

while a concrete WF for electron, say (4.3) with $|\boldsymbol{j}_D| > |\boldsymbol{c}_D|$ will transform into a WF for positron like (4.5) with $|\boldsymbol{c}_D^c| > |\boldsymbol{j}_D^c|$.

To see another hidden symmetry, we perform a linear transformation:

$$\boldsymbol{x}_D = \frac{1}{\sqrt{2}}(\boldsymbol{j}_D + \boldsymbol{c}_D)\ ,\quad \boldsymbol{h}_D = \frac{1}{\sqrt{2}}(\boldsymbol{j}_D - \boldsymbol{c}_D) \quad (7.8)$$

to recast (7.6) into:

$$\begin{cases} i\hbar \dot{\boldsymbol{x}}_D = ic\hbar \vec{\boldsymbol{s}} \cdot \nabla \boldsymbol{x}_D + m_o c^2 \boldsymbol{h}_D \\ i\hbar \dot{\boldsymbol{h}}_D = -ic\hbar \vec{\boldsymbol{s}} \cdot \nabla \boldsymbol{h}_D + m_o c^2 \boldsymbol{x}_D \end{cases} \quad (7.9)$$

which remains invariant under the pure space inversion ($x \to -x,\ t \to t$) with:

$$\boldsymbol{x}_D(-x,t) \to \boldsymbol{h}_D(x,t)\ ,\quad \boldsymbol{h}_D(-x,t) \to \boldsymbol{x}_D(x,t) \quad (7.10)$$

showing the parity conservation law. An electron's helicity may be left-handed if $|\boldsymbol{x}_D| > |\boldsymbol{h}_D|$ or right-handed if $|\boldsymbol{x}_D| < |\boldsymbol{h}_D|$.

The situation is drastically different for neutrino's equation (6.3), which is no longer invariant under the space-inversion transformation (7.10), showing the parity being violated to maximum----only $\boldsymbol{n}_L$ and $\bar{\boldsymbol{n}}_R$ exist whereas $\boldsymbol{n}_R$ and $\bar{\boldsymbol{n}}_L$ are forbidden. By contrast, Dirac equation is noninvariant under the pure time inversion, which is enjoyed by Eq. (6.3) as shown in (6.9). However, both (7.6) and (6.3) (in terms of $\boldsymbol{j}_i$ and $\boldsymbol{c}_i$) enjoy the same basic space-time inversion symmetry with transformation (7.7).

Two things need to be emphasized: First, as a simple linear transformation (7.8) can alter the meaning of WF drastically---- from hidden particle and antiparticle fields to hidden left and right handed chiral fields, we should not interpret the WF too materially.

For example, a neutrino has $|x_i| > |h_i|$, but $h_i$ cannot be explained as its ingredient of right-handed polarized state because it is in a 100% left-handed state (helicity = -1) explicitly. The paradox of "Schrödinger's cat" in QM is basically over[2].

Second, the analytical continuation of mass shown in (6.2) corresponds to substitution:

$$x_D \to x_i, \qquad h_D \to i h_i \qquad (7.11)$$

implying that in comparison with $x_D$ and $h_D$, $x_i$ and $h_i$ have an extra phase difference $90^0$, which makes neutrino a tachyon. We see again that the phase in WF is really the most subtle thing in QM.

## VIII. Divergence Difficulty in Quantum Field Theory

Since the quantum theory evolved from QM to the realm of Quantum field theory (QFT), various divergences have been bothering physicists a lot. [28]. It was thought that the divergence might stem from the "point-particle model" of QM. However, as discussed in previous sections, there is no point particle in QM. So the divergence has nothing to do with the "point-particle model" at all.

Beginning from a simple regularization-renormalization method (RRM) proposed by J.F.Yang (then a PhD candidate in 1994 [33]) , we have been getting rid of all divergences, counter terms, bare parameters and arbitrary running mass scale (after some other RRM) in QFT [14]. What we eventually realize is that we were overlooking again the principle of relativity stressed by Einstein. How can we calculate or even merely modify, say, an electron mass on its mass shell solely via finite loop evaluation? No, we can't. What we can understand is either no mass scale or two mass scales, but never one mass scale. This scenario was clearly shown in the Gross-Neveu model [34]. So the appearance of divergence is essentially a warning that we expected too much in doing calculation on perturbative QFT. What we can do is to substitute divergences by some arbitrary constants and then the renormalization amounts to fix these constants by experiments, say, to reconfirm the value of mass on its mass shell before we can predict what will happen in the off-mass-shell situation and go to the next higher loop calculation. As a metaphor, one has to reconfirm the plane ticket before his departure from the airport, he must use the same name through out his whole journey[14].

By adopting the new RRM, our calculation on QFT becomes more predictive and free of ambiguity. For instance, based on the standard model of particle physics and the accurate experimental data of masses of W and Z bosons as well as the value of Weinberg angle, we were able to predict the Higgs mass being around 138 GeV/$c^2$ ( not a lower or upper bound)[35], which should be tested experimentally in the years to come.

## IX. Summary and Discussion

(a). We have been learning from SR that the essence of the principle of relativity is nothing but the equal existence of particle and antiparticle. To be precise, every particle has two contradictory fields, $j$ and $c$, and the theory must obey the symmetry of space-time inversion like (7.7) with Eqs.(3.2) versus (4.4). This is the only one "relativistic postulate" (in one inertial frame) which would arouse the surprise of physicists living in 1905 to the same degree just like the postulate of constancy of light speed did.

(b). An alternative statement is: a theory , either quantum or classical, capable of treating particle and antiparticle on an equal footing should be invariant under the mass inversion m→ - m.

(c). In fact, the above two equivalent symmetries are combined in the Casimir invariant (2.6). Just like the four-dimensional space-time interval (2.3) can be positive, zero or negative, now in the four-dimensional momentum space:

$$E^2 - p^2c^2 = m_0^2 c^4 \qquad (9.1)$$

can also be positive, zero or negative. In the last case, as shown in (6.2), we will get a tachyonic neutrino. However, while $m_0 \to im_s$ is possible, so does $m_0 \to -im_s$, which means an antineutrino. Hence we see again that the mass inversion is indeed a symmetry transformation between i and – i ,which of course is a relative one , rather than an absolute one.

(d). In some sense, we may regard the two sets of operators, (3.2) and (4.4), as the DNA inherited by RQM, QFT and particle physics from QM and SR respectively[36,2].

(e).The principle of relativity is so deep and subtle that even Einstein himself had struggled for eight years before he invented the theory of general relativity on the basis of SR. What Einstein eventually realized is that space and time have no absolute meaning but systems of relations ([3],p.149).

We wish to add that the conception of time is essentially endowed by the observer (not by the object) as shown in the expression of WF (7.1) in QM. We build our concept of time t on experiences to observe a chain of events----from a cause to an effect, then a next effect, and on and on. So the time evolves always forward and never backward, a cognition which is totally in conformity with the positiveness  of mass and energy, also the fact that we are composed of Dirac particles violating the pure time inversion to maximum as discussed in section VII.  In other words, we always have a "arrow of time" not only on the macroscopic level as explicitly shown by the second law of thermodynamics , but also on the microscopic level implicitly as well.

(f). In 1935, Einstein, Podolsky and Rosen (EPR) wrote a seminal paper titled " Can quantum mechanical description of physical reality be considered complete?"[Phys. Rev. $\underline{47}$, 777, (1935)]  which has two far-reaching consequences.  First, its original version actually infers the necessity of introducing antiparticles to QM and fixes their operators as (4.4) [37, 2]. Second, equally importantly, it opened a fast developing field of physics---- the Bell's inequality [38] and its tests in various EPR experiments, especially that by Aspect et al[39], Weihs et al[40], Tittel et al[41] and the CPLEAR Collaboration[42] etc, showing the entanglement of quantum states. One common conclusion can be drawn from these experiments is: the concrete form of correlation in a quantum state exhibits itself as a direct outcome of measurement right at the destruction of original entanglement of state. If taking further the strong hints by the  remarkable "which-way" experiment [43] and SQUID experiment [44] into account, we may claim with confidence that the information does not exist in an object before the measurement, rather, it is created by subject and object in common during the measurement which is nothing but a changing process on the object.

(g). In short, we might summarize the above discussion by one English letter: "i" versus "I". The symmetry between i and –i implies the common essence of QM and SR, (without (4.4) provided by SR, QM cannot be regarded as complete). On the other hand, "I" means information, also represents the subject. Nonetheless, all descriptions on the objective world, including the theory of SR, the introduction of i and WF into QM etc, are all innovations of human being . After 1935, Einstein kept asking that "Is the moon there when nobody looks?"[45] It's time to answer this question now [2].